\documentclass[twoside]{dis07}
\usepackage[latin1]{inputenc}
\usepackage[dvips]{graphicx,epsfig,color}
\usepackage{wrapfig,rotating}
\usepackage{amssymb,amsmath,array}

\newcommand{\pom}{I\!\!P}
\begin{document}
\title{Charm production in diffractive DIS and PHP at ZEUS}

\author{Isabell-Alissandra Melzer-Pellmann, on behalf of the ZEUS collaboration
%
\vspace{.3cm}\\
%
 DESY, ZEUS group \\
Notkestr. 85, 22607 Hamburg, Germany
}

\maketitle

\begin{abstract}
The ZEUS experiment has measured charm production in diffractive DIS and 
in photoproduction. The data are in
agreement with perturbative QCD calculations based on various parameterisations
of diffractive parton distribution functions. The results are consistent with 
QCD factorisation in diffractive DIS and direct photoproduction \cite{url}.
\end{abstract}

\section{Introduction}

In diffractive electron-proton scattering,
the proton loses a small fraction of its energy and
either emerges from the scattering intact, $ep \to eXp$,
or dissociates into a low-mass state $N$, $ep \to eXN$.
Hadronic states $X$ including a $c\bar{c}$ pair are
a particularly interesting component of diffractive interactions.
The charm-quark mass provides a hard scale, 
ensuring the applicability of perturbative QCD even
for small photon virtualities (photoproduction).
At leading order (LO) of QCD, two types of  photoproduction processes
can be distinguished: direct and resolved photon processes.
Charm production mainly proceeds via direct photon reactions,
in which the exchanged photon participates as a point-like particle,
directly interacting with a gluon from the incoming proton
(photon-gluon fusion).
Thus, diffractive charm production is directly
sensitive to the gluon content of the diffractive exchange.
In resolved photon processes, the photon behaves
as a hadron-like source of partons, one of which interacts with
a parton from the initial proton.

\section{Diffractive D* in photoproduction}

ZEUS has recently measured diffractive $D^*$ in photoproduction \cite{ZeusDdPHP}
in the kinematic range $Q^2 < 1$ GeV$^2$, $130 < W < 300$ GeV, 
with transverse momentum $p_T(D^*) > 1.9$ GeV
and pseudorapidity\footnote{The ZEUS coordinate
system is a right-handed Cartesian system, with the $Z$ axis pointing
in the proton beam direction, referred to as the ``forward direction'',
and the $X$ axis pointing left towards the centre of HERA.
The coordinate origin is at the nominal interaction point.}
 $|\eta(D^*)|<1.6$, using an integrated 
luminosity of $\mathcal L=$78.6 pb$^{-1}$. Diffractive events were identified by 
a large gap in pseudorapidity between the produced hadronic state and 
the outgoing proton, $\eta_{\rm max}< 3$, reducing the fraction of proton 
momentum carried by the Pomeron to $x_{\pom} <0.035$. In addition,  
the energy deposited in the forward plug calorimeter (FPC), installed in the 
$20 \times 20$ cm$^2$ beam hole of the forward uranium calorimeter, 
was required to be consistent with zero ($E_{\rm FPC}<1.5$ GeV).
After all selections, a signal of $458 \pm 30$ $D^*$ mesons
was found. In order to reduce the contributions from 
non-diffractive background, the selection was also performed in the restriced
range $x_{\pom} <0.01$, where $204 \pm 20$ $D^*$ mesons were observed.
Proton-dissociative events can also satisfy the diffractive selection 
requirements if the proton-dissociative system, $N$,
has an invariant mass small enough to pass undetected through
the forward beam-pipe. The fraction ($f_{\rm pd}$) of these events was measured
previously to be $f_{\rm pd}=16 \pm 4(\rm syst.)$\%  \cite{DdDISpap}.

Cross section predictions for diffractive photoproduction of $D^*$ mesons
were calculated at the next-to-leading order (NLO) in  $\alpha_s$,
using the fixed-flavour-number scheme, in which only light flavours
are active in the parton distribution functions (PDFs) 
and the heavy quarks are generated by the hard
interaction.
The calculation was performed with the FMNR code
in the double-differential mode \cite{FMNR, applNLOQCD}.
The Weizs\"acker-Williams approximation \cite{WWA} was used
to obtain the virtual photon spectrum
for electroproduction with small photon virtualities.
Diffractive parton distribution functions (dPDFs) were used instead of the conventional
proton PDFs. The three sets of dPDFs used in the calculations were
derived from NLO QCD DGLAP fits to the HERA data on
diffractive deep inelastic scattering:
the H1 2006 Fit A, Fit B \cite{H1Fit2006} and
the ZEUS LPS+charm Fit \cite{ZEUSLPS} dPDFs.
In the ZEUS LPS+charm Fit, the diffractive DIS data were
combined with the results on diffractive
charm production in DIS \cite{DdDISpap} to better
constrain the gluon contribution.
The Reggeon contribution, which amounts to less than 2\%
for $x_{\pom}=0.01$ and grows up to $\sim 15$\%
at $x_{\pom}=0.035$,
was not included. To account for the proton-dissociative contribution,
present in the H1 2006 fits, the corresponding predictions
were multiplied by the factor 0.81 \cite{H1Fit2006}.
The calculations were performed with
$\alpha_s(M_Z)=0.118$ GeV \cite{PDG2006} and $m_c=1.45$ GeV,
the same values as used in the QCD fits to the HERA data.
The fraction of charm quarks hadronising as $D^*$ mesons
was set to $f(c \to D^*) = 0.238$ \cite{cFF}.
The Peterson parameterisation \cite{cFFPeter} was used for the charm
fragmentation with the Peterson parameter $\epsilon= 0.035$.
The uncertainties of the calculations were estimated by
varying the renormalisation and factorisation
scales simultaneously with the charm mass
to $\mu_R=\mu_F=0.5 \cdot \mu$, $m_c=1.25$ GeV and to
$\mu_R=\mu_F=2 \cdot \mu$, $m_c=1.65$ GeV.
Uncertainties on the dPDFs were not included.

The differential cross section for $ep \to eD^*X^{\prime}p$
in a given variable $\xi$ was calculated from
\[
\frac{d\sigma}{d\xi}=\frac{N_{D^*}\cdot(1-f_{\rm non-diffractive})\cdot(1-f_{\rm proton-dissociation})}
 {\mathcal{A} \cdot \mathcal{L}\cdot \mathcal{B} \cdot\Delta\xi}~,
 \]
where $N_{D^*}$~ is the number of
$D^*$ mesons observed in a bin of size $\Delta\xi$.
The overall acceptance was $\mathcal{A}=13.9\%$.
The combined \mbox{\( D^* \to (D^0 \to K \pi)\,\pi_s \)} decay branching ratio
is \( \mathcal{B}=0.0257 \pm 0.0005\) \cite{PDG2006}.
The measured cross sections are 
\begin{eqnarray*}
 \sigma_{ep\to  eD^* X^{\prime}p}(x_{\pom}<0.035 ) & = &
1.49 \pm 0.11(\rm stat.)^{+0.11}_{-0.19}(\rm syst.) \pm 0.07(p.d.)\; {\text nb},\; {\text and} \\
 \sigma_{ep\to eD^* X^{\prime}p}(x_{\pom}<0.01) & = &
0.63 \pm 0.07(\rm stat.)^{+0.04}_{-0.06}(\rm syst.) \pm 0.03(p.d.)\; {\text nb}.
\end{eqnarray*}
The last uncertainty is due to the subtraction of the
proton-dissociative background.
                                                                                
\begin{figure} [t]
\centerline{\includegraphics[width=0.4\textwidth]{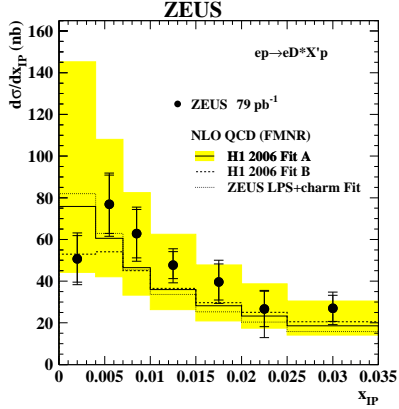}}
\caption{Differential cross section (dots) for diffractive photoproduction
of $D^*$, measured as function of $x_{\pom}$.
The inner bars show the statistical errors;
the outer bars correspond to the statistical
and systematic uncertainties added in quadrature.
The data are compared to the NLO QCD calculations (histograms).
The shaded bands show the uncertainties coming from variations
of the charm-quark mass and the
factorisation and renormalisation scales.}\label{Melzer:fig1}
\end{figure}

The differential cross section as function of $x_{\pom}$, shown in 
Fig.~\ref{Melzer:fig1}, demonstrates that the NLO predictions based on the 
various parameterisations of dPDFs are consistent with the data,
supporting the validity of diffractive QCD factorisation in diffractive 
direct photoproduction.
Differential cross sections as function of $M_X$, $p_T(D^*)$, $\eta(D^*)$, 
$z(D^*)$ and $W$ have also been calculated \cite{ZeusDdPHP}.

The ratio of the diffractive to the inclusive ($ep\to eD^*Y$)
photoproduction cross sections for $D^*$ mesons was also evaluated, 
as systematic uncertainties partly cancel in this ratio, which is defined as
$\mathcal{R_D}(D^*) =
\sigma_{ep\to eD^* X^{\prime}p}(x_{\pom}<0.035)/{\sigma_{ep\to eD^* Y}}$.
 
In the kinematic region $Q^2 < 1$ GeV$^2$,
$130 < W < 300$ GeV $(0.17<y<0.89)$,
$p_T(D^*)> 1.9$ GeV and $|\eta(D^*)| < 1.6$,
diffractive production for  $x_{\pom}< 0.035$ contributes 
\[ \mathcal{R_D}(D^*) = 5.7 \pm
0.5(\rm stat.)^{+0.4}_{-0.7}(\rm syst.)  \pm 0.3(p.d.)\%~ \]
to the inclusive $D^*$ photoproduction cross section.
The systematic uncertainty is dominated by the measurement of
the diffractive cross section.

\begin{figure} [!ht]
\centerline{\includegraphics[width=0.4\textwidth]{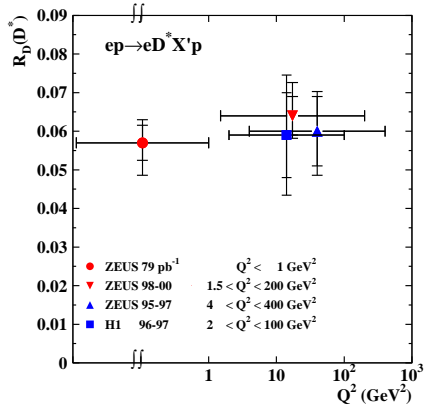}}
\caption{Fractions $\mathcal{R_D}$ of
$D^*$ meson diffractive production cross sections
measured at HERA in DIS {\cite{DdDISpap,ZeusDdDIS,DdDISH1}}
and photoproduction \cite{ZeusDdPHP}.
The inner bars show the statistical errors, the outer bars
correspond to the statistical and systematic uncertainties
added in quadrature.
\label{Melzer:fig2}}
\end{figure}

The NLO QCD predictions for $\mathcal{R_D}$ were obtained as the
ratio of the diffractive cross section, calculated
with the H1 2006 or ZEUS LPS+charm dPDFs, and the inclusive
cross section, obtained with the CTEQ5M proton PDFs.
The fraction $\mathcal{R_D}$ agrees with the values
measured at HERA for diffractive DIS in similar kinematic ranges
\cite{DdDISH1, DdDISpap, ZeusDdDIS}. As shown in Fig.~\ref{Melzer:fig2},
$\mathcal{R_D}$ shows no dependence on $Q^2$ within the errors of the 
measurement. 
Differential distributions as function of $p_T(D^*)$, $\eta(D^*)$, 
$z(D^*)$ and $W$ have also been calculated \cite{ZeusDdPHP}.
                                                                                
\section{Diffractive D* in DIS}

Diffractive $D^*$ in DIS has been measured by ZEUS \cite{DdDISpap}
in the kinematic range $1.5 < Q^2 < 200$ GeV$^2$, $0.02 < y < 0.7$,
$x_{\pom} <0.035$ and $\beta < 0.8$, where $\beta$ is the fraction of the Pomeron 
momentum carried by the struck quark. 
The transverse momentum of the $D^*$ is restricted to $p_T(D^*) > 1.5$ GeV
and its pseudorapidity to $|\eta(D^*)|<1.5$. Using an integrated 
luminosity of 82 pb$^{-1}$, $253 \pm 21$ $D^*$ candidates have been found. 
The data have been compared to the perturbative QCD
calculation program HVQDIS \cite{HVQDIS}, based on a parameterisation from a 
gluon dominated fit to H1 and ZEUS inclusive diffractive DIS and 
ZEUS diffractive photoproduction data \cite{ACTW}. The data are in good agreement
with the NLO predictions, confirming QCD
factorisation in diffractive DIS as proven by Collins \cite{Collins}.
The ratio $\mathcal{R_D}$ has been calculated as in the photoproduction
analysis and is also shown in Fig \ref{Melzer:fig2}.

\section{Conclusions}
Differential cross sections for diffractive $D^*$ production in photoproduction
and DIS have been compared to the predictions of NLO QCD calculations. 
The NLO predictions based on various 
parameterisations of diffractive PDFs are consistent with the data.
The measured fraction of $D^{*\pm}$ meson photoproduction
due to diffractive exchange is consistent with the measurements of $D^{*\pm}$ 
meson production in diffractive DIS.
Within the experimental uncertainties,
this fraction shows no dependence on $Q^2$ and $W$.
                                                                                
The results demonstrate that diffractive open-charm production 
is well described by the dPDF parameterisations extracted
from diffractive DIS data, supporting the validity
of diffractive QCD factorisation in diffractive DIS and direct photoproduction.
Given the large experimental and theoretical
uncertainties and the small hadron-like contribution expected
by the NLO calculations, no conclusion can be drawn for the
resolved photoproduction.


\begin{footnotesize}


\end{footnotesize}


\end{document}